\documentclass{nature}
\usepackage{feynmp}
\usepackage{amssymb}
\usepackage{amsmath}
\usepackage{epsfig}
\usepackage{array,graphicx,subfigure,slashed}
\usepackage{mathrsfs}
\usepackage{hyperref}
\usepackage{cases}
\usepackage{bbm}
\usepackage{bm}
\usepackage{gensymb}
\usepackage{float}
\usepackage{color}
\makeatletter
\let\saved@includegraphics\includegraphics
\AtBeginDocument{\let\includegraphics\saved@includegraphics}
\renewenvironment*{figure}{\@float{figure}}{\end@float}
\makeatother

\def\l{\left}
\def\r{\right}
\def\bea{\begin{eqnarray}} \def\eea{\end{eqnarray}}
\def\nn{\nonumber}

\def\be{\begin{equation}}       \def\ee{\end{equation}}
\def\bea{\begin{eqnarray}}      \def\eea{\end{eqnarray}}
\def\ba{\begin{array}}
\def\ea{\end{array}}
\def\bnum{\begin{enumerate} }
\def\enum{\end{enumerate}}
\def\l{\left}
\def\r{ \right}
\def\nn{\nonumber}

\def\=>{\Rightarrow}
\def\>{\rightarrow}

\def\Eq#1{Eq.~(\ref{#1})}
\def\Eqs#1{Eqs.~(\ref{#1})}
\def\Fig#1{Fig.~\ref{#1}}
\def\xk#1{\left(#1\right)}

\renewcommand{\v}[1]{{\bf #1}}

\renewcommand{\>}{\rangle}

\newcommand{\om}{\omega}

\title{Absence of emergent supersymmetry at superconducting quantum critical points
in Dirac and Weyl semimetals}

\author{Peng-Lu Zhao $^{1}$, Guo-Zhu Liu$^{1\ast}$}

\begin{document}

\maketitle

\begin{affiliations}
\item Department of Modern Physics, University of Science and Technology of
China, Hefei, Anhui 230026, China
\\
$^\ast$ Correspondence: Guo-Zhu Liu (gzliu@ustc.edu.cn)
\end{affiliations}

\begin{abstract}
Supersymmetry plays a crucial role in superstring theory and high
energy physics, but has never been observed in experiments.
Recently, an effective space-time supersymmetry was argued to emerge
in the low-energy region by tuning Dirac or Weyl semimetal to
approach a superconducting quantum critical point, at which the
Dirac or Weyl fermion and the bosonic order parameter are both
massless. Here, we study under what circumstances can space-time
supersymmetry be realized at a quantum critical point. We
demonstrate that the Yukawa-type coupling between the massless
fermion and massless boson can dynamically generate an infinite
number of non-supersymmetric terms in the effective field theory of
the boson. Owing to these terms, no space-time supersymmetry emerges
at the superconducting quantum critical points. The results provide
important constraint on the exploration of emergent space-time
supersymmetry in condensed matter systems.
\end{abstract}

\section*{INTRODUCTION}
Space-time supersymmetry (SUSY) is known to be the only nontrivial
combination of the space-time and internal symmetries
\cite{wessbook, Weinbergbook}. It transforms fermions into bosons,
and vice versa. SUSY provides a possible solution to hierarchy
problem \cite{wessbook, Weinbergbook, gervais71, wess74,
dimopoulos81}. Gravity is naturally obtained if SUSY is local in
space and time. Due to many fantastic features, SUSY has been
extensively studied in high energy physics for half a century.
Unfortunately, there is so far no experimental evidence for SUSY,
and it remains unclear whether or not SUSY is a fundamental law of
nature.

While the search for SUSY in high-energy processes has reached
basically nothing, a variety of condensed matter systems were
conjectured to manifest emergent SUSY at low energies
\cite{Friedan84, Nayak98, fendley03, Lee07, Huijse08, Yu10, Roy13,
Berg13, Grover14, Ponte14, Huijse14, Jian15, Zerf16, Li16, Li18,
Rahmani15, Maciejko16, Jian17}. At the microscopic level, these
systems are non-supersymmetric, and even non-relativistic. As the
energy is lowered down to zero, the systems can flow to a stable
fixed point at which the Lorentz symmetry is respected
\cite{Chadha83, Roy16}. Under certain circumstances, there might
emerge an effective space-time SUSY. The emergence of SUSY could
provide an opportunity to explore the intriguing predictions of SUSY
theories in condensed matter systems. However, it is technically
difficult to realize effective SUSY in actual materials, as one
usually needs to delicately tune two or even more parameters.

Quantum critical system is an ideal platform to probe emergent SUSY.
A prominent example is the quantum critical point (QCP) between
Dirac semimetal (SM) phase and uniform superconducting (SC) phase on
the surface of three-dimensional (3D) topological insulator
\cite{Grover14}. The Dirac fermion is massless (gapless) in the SM
phase due to the peculiar electronic structure, and becomes massive
(gapped) in the SC phase due to Cooper pairing. The SC order
parameter can be described by a composite boson, which is massive
(gapped) in the SM phase. As the system is tuned to exactly the QCP,
both the fermion and boson are strictly massless. Renormalization
group (RG) analysis \cite{Lee07, Grover14} argued that the model of
such a QCP flows to a stable infrared fixed point that respects an
effective space-time SUSY at low energies. An attractive feature of
this proposal is that SUSY is realized by tuning one single
parameter, which is technically more feasible than previously
studied models. SUSY was also predicted to emerge when a Dirac/Weyl
SM undergoes quantum phase transition to the pair-density-wave (PDW)
state \cite{Jian15, Jian17}. A universal characteristic of these
models is that the boson arises from some sort of fermion pairing.

Realizing space-time SUSY at large distances (low energies) in
quantum critical systems which are non-supersymmetric at small
distances is a fascinating notion. This might allow us to employ the
powerful methods developed in the studies of supersymmetric quantum
field theories \cite{wessbook, Weinbergbook} to investigate the
striking quantum critical phenomena. However, before ascertaining
the existence of emergent SUSY in any realistic quantum critical
system, we must first make sure that all the potential SUSY-breaking
effects are absent at low energies.

For effective SUSY to emerge, the low-energy field theory of QCP
must contain a number of massless fermions and an equal number of
bosons. The fermion-boson interaction is described by the
Yukawa-type coupling. From previous research experiences of
non-relativistic QCPs \cite{Chubukov03, Chubukov04,
ChubukovPepinRech04, Metlitski10, Metzner15} with dynamical
exponents $z=2$ and $z=3$, we have learned that the fermion-boson
coupling may lead to certain type of infrared singularity at
ultra-low energies in case both fermions and bosons are massless
(gapless). The infrared singularity may prevent SUSY from emerging
at low energies. For the Dirac and Weyl SMs considered in this work,
the dynamical exponent is $z=1$. We will perform a generic,
model-independent field-theoretic analysis of the Yukawa coupling
between massless fermion and massless boson, and demonstrate that
SUSY cannot emerge if the Yukawa coupling generates nonlocal
self-coupling terms of the boson field. Based on this result, we
establish a necessary condition for space-time SUSY to emerge at any
QCP. This condition is universally applicable and imposes an
important constraint to the exploration of emergent SUSY in
condensed matter systems.

\section*{RESULTS}

\textbf{Model}. While our analysis is entirely model independent,
here for concreteness we first consider a specific model to
illustrate our strategy. The generalization to other models is
straightforward. In particular, we choose to start from the
following action
\begin{eqnarray}
S = \int d\tau d^{d}r \Big[ i\bar{\psi} \slashed {\partial}\psi +
|\partial_{\mu}\phi|^2+\lambda_4|\phi|^{4} +
g(\phi^*\psi^{T}i\sigma_{2}\psi + \mathrm{h.c.})\Big],
\label{Eqaction}
\end{eqnarray}
where $\tau$ is imaginary time and $\slashed{\partial} =
\gamma_\mu\partial_\mu$. The $\gamma$-matrices are defined via the
Pauli matrices $\sigma_i$ as follows: $\gamma_0=\sigma_3$,
$\gamma_1=\sigma_1$, and $\gamma_2=\sigma_2$. This action can be
regarded as an effective theory for SM-SC QCP \cite{Zerf16}. The
boson field $\phi$ represents the SC order parameter, whereas the
spinor field $\psi$, whose conjugate is $\bar{\psi} = -i\psi^\dag
\gamma_0$, stands for the Dirac fermion. According to RG study
\cite{Zerf16}, fermion and boson velocities can be taken to be
equal, i.e., $v_f = v_b$. The strength parameter for Yukawa coupling
is $g$, and that for $|\phi|^4$ coupling is $\lambda_4$. When the
system is deep in the SM phase, there are only fermionic degrees of
freedom, namely Dirac fermion. As the system is tuned to approach
the QCP towards SC order, composite boson emerges due to Cooper
pairing. At the QCP, both the fermion and the boson are exactly
massless and equally important at low energies. The Yukawa coupling
between them determines the quantum critical behaviors
\cite{Chubukov03, Chubukov04, ChubukovPepinRech04}. Three-loop RG
analysis performed by Zerf \emph{et al.} \cite{Zerf16} indicates
that this model has a stable infrared fixed point $g^{\ast2}=
\lambda_4^{\ast}$, at which the action (\ref{Eqaction}) is invariant
under the following SUSY transformations
\begin{eqnarray}
&&\delta_\eta\psi = i\slashed{\partial}\phi^*\eta -
\frac{1}{2}\phi^2i\sigma_2\bar{\eta}^T, \quad \delta_\eta\phi =
-\bar{\psi}\eta, \\
&&\delta_\eta\bar{\psi} = i\bar{\eta}\slashed{\partial}\phi -
\frac{1}{2}\phi^{*2}\eta^T i\sigma_2, \quad \delta_\eta\phi^* =
\bar{\eta}\psi,
\end{eqnarray}
where $\eta$ is a two-component Grassmann variable introduced to
define the transformations.

\begin{figure}[htbp]
\centering
\subfigure[]{\includegraphics[width=1.6in]{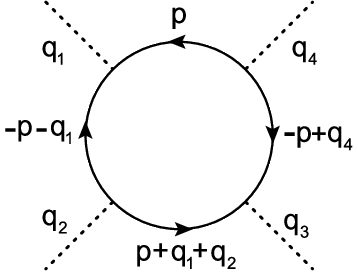}\label{Figb2nDW1}}
\hspace*{1.5cm}
\subfigure[]{\includegraphics[width=1.6in]{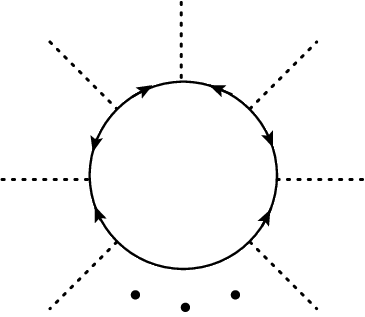}\label{Figb2nDW2}}
\caption{Feynman diagrams for the vertex corrections. (a) Diagram
for the vertex correction to $|\phi|^4$ term. (b) Diagram for the
vertex correction to a generic $|\phi|^{2n}$ term with $n>2$. The
solid (dashed) line represents the free fermion (boson) propagator.}
\label{Figb2nDW}
\end{figure}

\textbf{Infrared singularities}. To examine whether SUSY emerges, it
is necessary to carry out perturbative RG analysis of
Eq.~(\ref{Eqaction}). Although the boson originates from fermion
pairing, the coupling constants $\lambda_4$ and $g$ could flow
independently with varying energy scale. In the low energy limit, if
the system flows to a stable infrared fixed point
$(g^{\ast},\lambda_4^{\ast})$ satisfying the condition $g^{\ast2}=
\lambda_4^{\ast}$, SUSY emerges. This is the strategy adopted in
previous studies on emergent SUSY \cite{Lee07, Grover14, Zerf16}.
However, all the previous RG studies are based on an assumption that
the $|\phi|^4$ term is always a smooth function of momentum vector
$q=(q_0,\mathbf{q})$. This assumption is valid only when the
$|\phi|^4$ term does not receive any singular quantum correction
from the Yukawa coupling. In Fig.~\ref{Figb2nDW1}, we show the
one-loop diagram for the correction to $|\phi|^4$ term. In case the
contribution of this diagram is singular (non-analytic) in $q$, the
low-energy properties of the $\lambda_4|\phi|^4$ term would be
dominated by the singular correction, and the total $|\phi|^4$ term
would become nonlocal. The nonlocal $|\phi|^4$ term does not respect
SUSY.

Apart from $|\phi|^4$ term, the Yukawa coupling also produces an
infinite number of higher order terms $|\phi|^{2n}$, where the
integer $n > 2$. The corresponding diagram is given by
Fig.~\ref{Figb2nDW2}. In previous works on emergent SUSY
\cite{Lee07, Grover14, Jian15, Zerf16, Jian17}, all of such
$|\phi|^{2n}$ terms were naively supposed to be irrelevant and thus
were simply ignored. But there is no guarantee that these terms are
irrelevant at QCP. Indeed, we will show below that all of these
terms are marginal at low energies. The key point is that, the
Yukawa coupling induces singular corrections to $\lambda_4$ and as
such alters the scaling dimension of the boson field $\phi$. Once
this fact is taken into account, all the $|\phi|^{2n}$ terms with
$n>2$ are important at low energies.

We infer from the above analysis that SUSY emerges only when the
following condition is fulfilled: the Yukawa coupling does not lead
to any nonlocal $|\phi|^{2n}$ term. If one or more $|\phi|^{2n}$
terms are generated, SUSY is absent.

In the following, we calculate the coefficients of $|\phi|^{2n}$
terms with $n \geq 1$, and then make a scaling analysis to verify
whether these SUSY-breaking terms are relevant, marginal, or
irrelevant. Our analysis starts from the local action given by
Eq.~(\ref{Eqaction}), which is formally similar to, but not the same
as, the models studied in Refs.~\cite{Lee07, Grover14, Ponte14,
Jian15, Jian17}. It is straightforward to apply our
model-independent analysis to models defined at any dimension and
made out of any type of fermion (Dirac or Weyl).

As the first step, we compute the diagrams Fig.~\ref{Figb2nDW1} and
\ref{Figb2nDW2}. The loop integral contains a number of free fermion
propagators. When the fermion is massive (gapped), the integral is
free of infrared singularity since the mass provides an infrared
cutoff that regularizes the small energy/momenta contributions. But
since the fermion is massless at the QCP, the integral might be
severely divergent in the infrared region. One should compute the
loop integral carefully. According to action (\ref{Eqaction}), the
free fermion propagator is $G_0(k) = 1/\slashed{k}$. The vertex
correction presented in Fig.~\ref{Figb2nDW1} is defined as
\begin{eqnarray}
\lambda_4(q_1,q_2,q_3) &=&g^4\int \frac{d^Dp}{(2\pi)^D}
\mathrm{Tr}\Big[G_0(p)\sigma_2 G_0^T(-p-q_1) \sigma_2
G_0(p+q_1+q_2)\sigma_2 \nonumber \\
&& \times G_0^T(-p-q_1-q_2-q_3)\sigma_2\Big],\label{Eqlambda4}
\end{eqnarray}
where the space-time dimension $D = d+1$ and we have used the fact
that $q_{4} = -q_1 - q_2 - q_3$. The vertex function depends on
three independent external momenta, namely $q_1$, $q_2$, and $q_3$.
It is difficult to get an analytic expression. Fortunately, the
possible singularity comes mainly from very small external momenta
\cite{Chubukov03, Chubukov04}. This fact allows us to assume that
$q_1 = q_2 = q_3 = q$. Carrying out analytical calculations, we
obtain
\begin{eqnarray}
\lambda_4(q,q,q) \propto \frac{g^4}{|q|}\label{Eqla42d}
\end{eqnarray}
in $D=2+1$ with a positive constant prefactor and $|q| =
\sqrt{q_0^2+\mathbf{q}^2}$, and
\begin{eqnarray}
\lambda_4(q,q,q)\propto g^4\ln \l(\frac{|q|}{\Lambda}\r)
\label{Eqla43d}
\end{eqnarray}
in $D=3+1$, where $\Lambda$ is an ultraviolet cutoff. Apparently,
$\lambda_4(q,q,q)$ is singular as $q \rightarrow 0$.

The loop diagram Fig.~\ref{Figb2nDW1} was calculated in previous
works \cite{Lee07, Grover14, Jian15, Zerf16} by means of standard RG
method. Although the models considered in these works are physically
different, the loop integrals have the same structure and can be
computed universally. No singular contribution was found in the
calculations of Refs.~\cite{Lee07, Grover14, Jian15, Zerf16}. That
no singular behavior was obtained in their works originates from the
treatment that all the external boson momenta were directly set to
zero \cite{Lee07, Grover14, Jian15, Zerf16}. We emphasize here that
this treatment is questionable. Notice that the dominant
contributions to diagram Fig.~\ref{Figb2nDW1} come from the
processes in which the internal fermion momenta $p$ are smaller than
the external boson momenta $q$, i.e., processes with $|p| < |q|$. If
all the external boson momenta are naively forced to vanish, these
contributions would be entirely discarded. In order not to miss the
contributions from the fermionic modes carrying ultra-low momenta,
the external boson momenta should be kept finite. As shown in
Eq.~(\ref{Eqla42d}) and Eq.~(\ref{Eqla43d}), the coupling parameter
$\lambda_4$ becomes singular once such contributions are carefully
taken into account, which makes the $|\phi|^4$ term nonlocal.

Numerical simulations \cite{Li16, Li18} have been performed to
calculate correlation functions at some SM-SC QCPs, claiming to find
evidence of emergent SUSY. The particle momenta obtained in
numerical simulations are discrete, thus a large part of the
ultra-low energy/momenta region cannot be accessed in numerical
simulations. This indicates that the SUSY-breaking nonlocal terms
are actually invisible in numerical data. Therefore, although
numerical simulations could obtain interesting information on the
correlation functions at intermediate energies, their resolution is
not high enough to unambiguously determine whether an exact
space-time SUSY is respected in the ultra-low energy region.
Moreover, in the strict sense SUSY is an extension of the
Poincar\'{e} space-time symmetry \cite{wessbook, Weinbergbook}. When
a model system is defined on finite lattices \cite{Li16, Li18}, both
the continuous translational symmetry and Lorentz symmetry are
broken. To judge whether exact space-time SUSY emerges in the lowest
energy limit, it is necessary to directly study continuum quantum
field theories and carry out careful field-theoretic analysis.

Previous studies have demonstrated \cite{Chubukov03, Chubukov04,
ChubukovPepinRech04, Metlitski10, Metzner15} that the quantum
critical fluctuations of order parameters may induce nonlocal terms
in some non-relativistic metals with dynamical exponent larger than
unity ($z>1$). This behavior is related to the existence of an
infinite number of gapless fermionic excitations on the finite Fermi
surface. For instance, such a coupling produces highly nonlocal
terms at the QCP between $d=2$ normal metal and antiferromagnet
\cite{Chubukov04}. Different from these examples, the fermion
density of states (DOS) vanishes at the Dirac/Weyl point of the
SM-SC QCP studied in this work. However, the vanishing of fermion
DOS cannot be used to preclude nonlocal terms. The reason is that
the particle momenta $q$ cannot be always fixed at zero: $q$ should
be allowed to take both zero and nonzero values. No matter how small
the boson momenta $q$ is, there are always an infinite number of
fermionic modes below the energy scale of $|q|$, which inevitably
gives rise to the infrared singularity of $\lambda_4(q)$.

After showing that $|\phi|^4$ term becomes nonlocal due to Yukawa
coupling, we next consider the $|\phi|^{2n}$ terms with $n > 2$. For
this purpose, we need to compute the diagram Fig.~\ref{Figb2nDW2}.
The coupling parameter for $|\phi|^{2n}$ terms is formally given by
\begin{eqnarray}
\lambda_{2n} &=&
g^{2n}\int\frac{d^Dp}{(2\pi)^D}\mathrm{Tr}\Bigg[\sigma_2
G_0(p)\sigma_2 G_0^T(-p-q_1)\sigma_2...
\nonumber \\&& \times
G_0\l(p+\sum_{i=1}^{2n-2}q_i\r)\sigma_2
G_0^T\l(-p-\sum_{i=1}^{2n-1}q_i\r)\Bigg]\label{Eqlambda2n},
\end{eqnarray}
where for notational simplicity the explicit dependence of
$\lambda_{2n}$ on external momenta is omitted. We adopt the same
simplification as what we have done in the computation of
$\lambda_4$, and set all the external momenta to be equal. Under
this approximation, we obtain \bea \lambda_{2n}\propto
g^{2n}\l(\frac{1}{q^2}\r)^{n-\frac{D}{2}}. \label{Equ2n} \eea
Strictly speaking, this expression is applicable only for $n\neq
D/2$. However, since our aim is to determine the scaling behavior of
the self-coupling vertices, it can be written in this form for any
$n$ and $D$. Detailed calculation of $\lambda_{2n}$ is presented in
the Supplementary Material. It is easy to observe that the coupling
parameter $\lambda_{2n}$ also becomes singular, and that the
singularity is more severe for larger $n$.

\textbf{Importance of nonlocal terms}. On the basis of
$\lambda_4(q)$ and $\lambda_{2n}(q)$, we next make a scaling
analysis to determine whether the non-local $|\phi|^{2n}$ terms are
relevant, marginal, or irrelevant. Incorporating the generated
nonlocal terms into the original action (1), we now re-write the
effective total action as follows
\begin{eqnarray}
S &=&  S_{f}^0 + S_{b}^0 + S_I + S_{b}, \label{EqactionBf} \\
S_{f}^0 &=& \int \frac{d^{D}p}{(2\pi)^D}\bar{\psi}(p)\gamma \cdot
p \psi(p), \\
S_{b}^0 &=& \int\frac{d^{D}k}{(2\pi)^D}k^2 \phi^{\dag}(k)\phi(k),
\\
S_I &=& \int\frac{d^{D}p}{(2\pi)^D}\frac{d^{D}k}{(2\pi)^D}
g\left[\psi^{T}(p)i\sigma_2 \psi(k)\phi(k-p) + h.c.\right], \\
S_{b} &=& \sum_{n=1}^{\infty}u_{2n} \l(\int \frac{d^D
k}{(2\pi)^{D}}\r)^{2n-1}(k^2)^{\frac{D}{2}-n}|\phi\l(k\r)|^{2n},
\label{EqSb}
\end{eqnarray}
where $u_{2n} \propto g^{2n}/v_f^d$. Using the scaling dimension in
momentum space, we have
\begin{eqnarray}
\l[p_0\r]=z,\quad \l[\mathbf{p}\r]=1,\quad \l[v_f\r]
=\l[v_b\r]=z-1,\quad \l[p\r] = \l[\sqrt{p_0^2 + v_{f,b}^2
\mathbf{p}^2}\,\,\r] = z.
\end{eqnarray}
Here, $z$ is the dynamical exponent for both the fermionic and
bosonic fields.

Within the standard RG framework \cite{Shankar94}, one usually takes
the free action $S^0 = S_{f}^0 + S_{b}^0$ as the free fixed point,
which reproduces itself upon making RG transformations, and then
determines the scaling behaviors of all the field operators and
coupling constants. In the present case, it is easy to find the
following scaling behaviors:
\begin{eqnarray}
&&\l[\psi(p)\r] = -\frac{2z+d}{2},\quad
\l[\phi(k)\r] =-\frac{3z+d}{2},  \nn\\
&&\l[g\r] = \frac{3z-d}{2},\quad
\l[u_{2n}\r] = (3z-d)n+d(1-z).\label{Eqscalfree}
\end{eqnarray}
According to Eq.~(\ref{Equ2n}), for $n = 1$ we see that the one-loop
polarization function behaves as \bea \Pi_2(k)\propto
g^2\l(k^2\r)^{\frac{D}{2}-1},\label{Eqpolarization} \eea which means
that $z = 1$ is not changed by adding the polarization to the free
boson action. As pointed out in Refs.~\cite{Kim08, Huh08}, that $z$
is not altered by higher order corrections results from the $U(1)$
gauge invariance \cite{Zerf16} of action (\ref{Eqaction}). In the
case of $z = 1$, it is clear that $\l[u_{2n}\r] = 0$ for $d = 3$,
thus all the $|\phi|^{2n}$ terms are marginal. For $d < 3$,
$\l[u_{2n}\r]>0$ and there are an infinite number of relevant
self-coupling terms. This indicates that, at the QCP of $d < 3$
system, the one-loop polarization $\Pi_2(k)$ is more important than
the free boson action given by Eq.~(\ref{EqactionBf}) in the
ultralow energy region. As a result, it is no longer appropriate to
regard the free boson action $S_b^0$ as the free fixed point. A more
suitable approach is to adopt the scheme proposed in
Refs.~\cite{Huh08, Sachdev}: discard the free boson action $S_b^0$,
and determine the scaling dimension of boson field $\phi$ by taking
the Yukawa coupling, i.e., $S_I$, as the starting fixed point.
Employing this scheme, we find the following correct scaling
dimensions:
\begin{eqnarray}
\l[\psi(p)\r] = -\frac{2z+d}{2},\quad \l[\phi(k)\r] = -d, \quad
\l[g\r] = 0,\quad \l[u _{2n}\r] = d(1-z).\label{Eqscalyukawa}
\end{eqnarray}
For $z = 1$, all the $|\phi|^{2n}$ terms with $n \geq 2$ are
marginal. This conclusion is independent of the spatial dimension.
An immediate indication is that all the bosonic self-coupling terms
are equally important at low energies and should be considered
simultaneously.

We then discuss the impact of nonlocal $|\phi|^4$ and $|\phi|^{2n}$
terms on the fate of emergent SUSY. In the absence of nonlocal
terms, the action given by Eq.~(\ref{Eqaction}) is a well-defined
local quantum field theory. All the $|\phi|^{2n}$ terms with $n>2$
are irrelevant, and can be simply ignored. RG calculations
\cite{Zerf16} confirm that the system flows to a stable infrared
fixed point at which an effective SUSY emerges. This conclusion,
however, is fundamentally changed once nonlocal terms are generated.
Since all the $|\phi|^{2n}$ terms with $n \geq 2$ are marginal, the
simple action (\ref{Eqaction}) is not the correct low-energy theory
of SM-SC QCP. Actually, after incorporating the marginal nonlocal
terms into action (\ref{Eqaction}), the SM-SC QCP does not exhibit
emergent SUSY.

\section*{DISCUSSION}

In supersymmetric field theories, there is a non-renormalization
theorem \cite{wessbook, Weinbergbook}, which ensures that loop
integrals are free of divergences. This theorem cannot be used to
eliminate the infrared singularities induced by the Yukawa coupling
in quantum critical systems. This is because the non-renormalization
theorem is applicable only in the special case in which the system
respects the intrinsic SUSY at the most microscopic level. However,
emergent SUSY refers to an effective SUSY that is realized only at
very low energies but is apparently absent at high energies (small
distances). To avoid infrared singularities, one could construct a
lattice model that respects space-time SUSY at the tree (classic)
level. Nonlocal terms would not exist in such a model due to the
non-renormalization theorem. To realize this type of lattice model,
one has to delicately tune two or more model parameters \cite{Yu10},
which is certainly difficult.

Now suppose the boson appearing in Eq.~(\ref{Eqaction}) is not made
out of fermion pairing. This sort of boson has its own dynamics. Can
SUSY emerge at large distances in such a system that is
non-supersymmetric at small distances? The answer is negative, since
the Yukawa coupling still induces nonlocal quantum correction to
each self-coupling term $|\phi|^{2n}$ with $n \geq 2$. The system
does not respect SUSY at the microscopic level, thus one cannot
invoke the non-renormalization theorem to prevent nonlocal
corrections. The nonlocal corrections dominate over the original
local terms at low energies, and all the $|\phi|^{2n}$ terms are
marginal. We therefore conclude that nonlocal terms prevent SUSY no
matter whether or not the boson arises from fermion pairing.

Our consideration is general for continuous quantum phase
transitions and independent of the specific model to achieve the
transition. As a concrete example, we have showed that the model
described action (\ref{Eqaction}) does not exhibit emergent SUSY due
to nonlocal $|\phi|^{2n}$ terms. The condition can be easily
extended to other similar models. Here, we mention two kinds of
models. The first kind is the SM-SC QCP realized on the surface of
3D topological insulator \cite{Grover14, Ponte14, Zerf16}. This QCP
can be described by the effective action (\ref{Eqaction}). The
second kind is defined by the QCP between Dirac/Weyl SM and PDW
\cite{Jian15}, which involves two Dirac/Weyl points. In this case,
the effective action can be regarded as two copies of
Eq.~(\ref{Eqaction}) along with an additional coupling between two
types of bosons. Although the PDW order parameter carries a finite
momentum \cite{berg2007, berg2009}, we confirmed that the Yukawa
coupling also induces nonlocal terms at the SM-PDW QCP. Thus, no
SUSY emerges in the above two kinds of models.

In conclusion, we have revealed a necessary condition for space-time
SUSY to emerge in a microscopically non-supersymmetric system: no
nonlocal self-coupling terms of the boson field are induced by the
Yukawa coupling. This provides an important guidance to the search
of emergent SUSY in condensed-matter systems. Because the Yukawa
coupling between massless fermion and massless boson often leads to
nonlocal terms, it is not legitimate to seek emergent SUSY at QCPs.
Instead, one should consider fully gapped systems. For instance, one
might try to tune the system away from the QCP to enter into the
ordered phase such that both the fermion mass $m_f$ and the boson
mass $m_b$ are finite \cite{Grover14}. If the distance to QCP took
certain suitable value, the two masses could be equal \cite{Li16},
namely $m_f = m_b$. In such a special case, nonlocal terms are
absent and in principle there might be effective SUSY. However, this
equal-mass state is fragile: once the distance to QCP is changed,
there is no chance for SUSY to emerge.

Experiments can be performed to judge whether SUSY is respected. In
case the system described by action Eq.~(\ref{Eqaction}) respects
space-time SUSY, the anomalous dimensions for fermion and boson
fields are equal and given by \cite{Aharony1997}
$\eta_{\psi}=\eta_{\phi}=1/3$. As a result, the fermionic DOS
behaves as $\rho\xk{\om}\propto \om^{4/3}$ at the QCP. In addition,
SUSY would lead to an exact result for the zero-temperature optical
conductivity \cite{Maciejko16} $\sigma\xk{\om,T=0} =
\frac{5\xk{16\pi - 9\sqrt{3}}}{243\pi}\frac{e^2}{\hbar}$. These two
quantities can be measured in experiments. Since our analysis found
that nonlocal terms prevent space-time SUSY from emerging, we
predict that the above signatures of SUSY would not be observed.
Recent experiments have reported the discovery of intrinsic
superconductivity on the surface of a 3D topological insulator
Sb$_2$Te$_3$\cite{LZhao16}, which allows one to experimentally test
whether SUSY really emerges at the SM-SC QCP.

The physical impact of nonlocal $|\phi|^{2n}$ terms is to be fully
understood. Such terms play a dominant role in the low-energy region
since their prefactors diverge as the lowest energy limit is taken.
The QCP becomes unstable due to these terms. Such an instability
could be avoided by two possible scenarios. Firstly, the phase
transition may become first order \cite{Belitz99}. At a first order
transition, the boson field $\phi$ develops a finite vacuum
expectation value abruptly, rather than continuously. In this case,
the boson mass $m_b$ is always finite and there are no nonlocal
terms. This scenario can be experimentally explored by measuring
latent heat and change of specific volume. Secondly, the quantum
fluctuation of bosonic order parameter may induce a new type of
ordered phase, which substitutes the putative unstable QCP at low
temperatures. At sufficiently high temperatures, the new order is
destroyed, and the Yukawa coupling does not generate nonlocal terms
but leads to intriguing quantum critical phenomena characterized by
finite anomalous dimensions of fermion and boson fields
\cite{Grover14}. This picture is analogous to the one in which the
strong nematic quantum critical fluctuation around a nematic QCP
induces long-range SC order at low temperatures and non-Fermi-liquid
behavior at high temperatures \cite{Metlitski15, Lederer15}. For
both of the two scenarios, no space-time SUSY emerges at the
transition. The model independent analysis presented in this work is
not sufficient to judge which scenario is the actual one, because
the answer might sensitively depend on the spatial dimensionality,
the nature of the fermion field, and the property of the bosonic
order parameter. Further investigations are called for to determine
which of these two scenarios works at a given QCP.

\section*{METHODS}

The nonlocal $|\phi|^{2n}$ terms are obtained by calculating the
Feynman diagrams presented in \Fig{Figb2nDW} using standard methods
of quantum field theory. In order not to miss such nonlocal terms,
the external boson momenta are kept finite. The calculational
details of \Eq{Eqlambda4} and \Eq{Eqlambda2n} can be found in the
Supplementary Material. The (ir)relevance of nonlocal $|\phi|^{2n}$
terms in the low-energy region is investigated by making a scaling
analysis of the effective action given by
\Eqs{EqactionBf}--(\ref{EqSb}). The scaling dimensions of the
fermion field, the boson field, and all the coupling constants
(i.e., $g$ and $u_{2n}$) are determined by taking the Yukawa
coupling term, rather than the free action, as the free fixed point.

\section*{DATA AVAILABILITY}
The authors declare that the data supporting the findings of this
study are available within the paper and its supplementary material.

\section*{ACKNOWLEDGEMENTS}
The authors thank X. Li, J.-R. Wang, J. Wang, and Y.-H. Wu for
helpful discussions, and acknowledge the financial support by the
National Natural Science Foundation of China under Grant 11574285.

\section*{AUTHOR CONTRIBUTIONS}
G.-Z.L. conceived the project and wrote the paper. P.-L.Z. carried
out the calculations.

\section*{ADDITIONAL INFORMATION}

\textbf{Supplementary information} accompanies the paper.\\
\textbf{Competing interests:} The authors declare no competing
interests.

%\textbf{Publisher's note:} Springer Nature remains neutral with
%regard to jurisdictional claims in published maps and institutional
%affiliations.

\newpage

\section*{Supplementary Material}

After including all the possible higher order self-coupling terms,
the boson action can be written in the generic form
\begin{eqnarray}
S_{b} &=& \int\frac{d^Dk}{(2\pi)^{D}} \l[\phi^{\dag}\l(k\r)D^{-1}
\l(k\r)\phi\l(k\r)+\sum_{n=2}^{\infty}
\l(\int\frac{d^Dk}{(2\pi)^{D}}\r)^{2n-1}
\lambda_{2n}|\phi\l(k\r)|^{2n}\r],\label{EqactionB}
\end{eqnarray}
where $D^{-1}(k) = D_0^{-1}(k)+\Pi(k)$ with $D_0^{-1}(k) = k^2$.

\begin{figure}[htbp]
\centering
\subfigure[]{\includegraphics[width=1.5in]{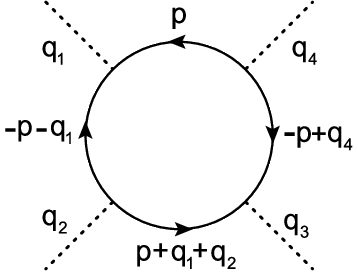}\label{Figb2nDW1A}}
\hspace*{1.5cm}
\subfigure[]{\includegraphics[width=1.5in]{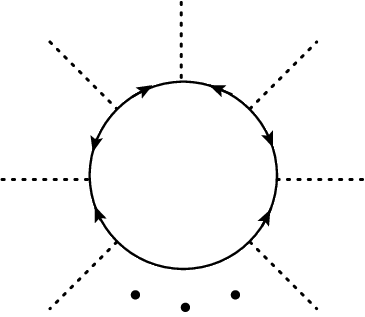}\label{Figb2nDW2A}}
\caption{(a) Feynman diagram for one-loop correction to $|\phi|^4$
term. (b) Feynman diagram for the quantum correction to a generic
$|\phi|^{2n}$ term with $n>2$. The solid (dashed) line represents
the free fermion (boson) propagator.} \label{Figb2nDWA}
\end{figure}

\section*{Nonlocal contributions to $|\phi|^4$ term}

The diagram of Fig.~\ref{Figb2nDW1A} is calculated as follows:
\begin{eqnarray}
\lambda_4(q_1^{\mu},q_2^{\mu},q_3^{\mu})\!\! \!\!&=&\!\!\!\!
g^4\!\!\int\!\! \frac{d^Dp}{(2\pi)^D} \mathrm{Tr}\Big[G_0(p)\sigma_2
G_0^T(-p-q_1) \sigma_2 G_0(p+q_1+q_2)\sigma_2
G_0^T(-p-q_1-q_2-q_3)\sigma_2\Big]
\nn\\
\!\! \!\!&=&\!\!\!\!  g^4\int
\frac{d^Dp}{(2\pi)^D}\mathrm{Tr}\Big[G_0(p)
G_0(p+q_1)G_0(p+q_1+q_2)G_0(p+q_1+q_2+q_3)\Big]
\nn\\
\!\! \!\!&=&\!\!\!\! 2g^4\int
\frac{d^{D}p}{(2\pi)^D}\frac{1}{p^2(p+q_1)^2
(p+q_1+q_2)^2(p+q_1+q_2+q_3)^2} \nn\\ &&\times
\Big\{\big[p\cdot(p+q_1)\big]
\big[(p+q_1+q_2)\cdot\l(p+q_1+q_2+q_3\r)\big] \nn \\
&& -\big[p\cdot(p+q_1+q_2)\big]
\big[(p+q_1)\cdot\l(p+q_1+q_2+q_3\r)\big] \nn \\
&& +\big[p\cdot\l(p+q_1+q_2+q_3\r)\big]
\big[\l(p+q_1\r)\cdot\l(p+q_1+q_2\r)\big]\Big\}.
\end{eqnarray}
In the above derivation, we have used the relation $\sigma^y
G_0^T(p)\sigma^y = G_0(-p)$. In general, it is hard to evaluate the
above integration in the presence of three freely varying outline
momenta. To simplify analytical calculation, we assume that
$q_1^{\mu} = q_2^{\mu} = q_3^{\mu}=q^{\mu}$. After tedious but
straightforward computations, we find that
\begin{eqnarray}
\lambda_4(q^{\mu},q^{\mu},q^{\mu}) = \frac{\l(20\sqrt{33} +
423\r)g^4}{6144}\frac{1}{|q|},
\end{eqnarray}
for $D=2+1$, where $|q|= \sqrt{q^{\mu}q^{\mu}} = \sqrt{q_0^2 +
\mathbf{q}^2}$.

For $D=3+1$, the above integral contains a logarithmic divergence in
the UV region. As we focus on the IR singularity, this divergence
can be controlled by introducing a UV cutoff $\Lambda$. It is easy
to obtain
\begin{eqnarray}
\lambda_4(q^{\mu},q^{\mu},q^{\mu}) = -\frac{g^4}{4\pi^2}
\ln\l(|q|^2/\Lambda^2\r),
\end{eqnarray}
where $q^2 = q_0^2 + \mathbf{q}^2$ and we have absorbed all the
constants into the re-scaled $\Lambda$.

\section*{General nonlocal $|\phi|^{2n}$ terms}

We then compute the diagram of Fig.~\ref{Figb2nDW2A}, which is given
by \bea \lambda_{2n}(q^{\mu},...,q^{\mu}) &=&
g^{2n}\int\frac{d^Dp}{(2\pi)^D} \mathrm{Tr}\Bigg[G_0(p)\sigma_2
G_0^T(-p-q_1)\sigma_2... G_0\l(p+\sum_{i=1}^{2n-2}q_i\r)\sigma_2
\nn\\ &&\times G_0^T\l(-p-\sum_{i=1}^{2n-1}q_i\r)\sigma_2\Bigg]
\nn\\&=& g^{2n}\int\frac{d^Dp}{(2\pi)^D} \mathrm{Tr}\l[G_0(p)
G_0(p+q_1)...G_0\l(p+\sum_{i=1}^{2n-1}q_i\r)\r] \nn\\&=&
2g^{2n}\int\frac{d^Dp}{(2\pi)^D}\frac{1}{\prod_{i=0}^{2n-1}(p+a_i)^2}
\times \Bigg[\sum_{i_1 = 1}^{2n-1}
(-1)^{r(a_{i_1},a_{i_2}...a_{i_{2n-1}})} p\cdot(p+a_{i_1})\nn\\
&&\times (p+a_{i_2})\cdot(p+a_{i_3})
\times...\times(p+a_{i_{2n-2}})\cdot(p+a_{i_{2n-1}})\Bigg], \eea
where $a_i = \sum_{i=1}^{i}q_i$, $(a_{i_1},a_{i_2}...a_{i_{2n-1}})$
is an arbitrarily array for $(a_1,a_2...a_{2n-1})$ and
$r(a_{i_1},a_{i_2}...a_{i_{2n-1}})$ is the total replacement number
of the array $(a_{i_1},a_{i_2}...a_{i_{2n-1}})$ with a condition
that $i_j < i_k$ is satisfied in every inner product term
$(p+a_{i_j})\cdot(p+a_{i_k})$. We set $q_1^{\mu} = q_2^{\mu} = ... =
q_{2n-1}^{\mu} = q^{\mu}$ and then introduce $(2n-1)$ parameters,
namely $x_1,x_2,...,x_{2n-1}$, to carry out the Feynman
parametrization. We eventually get \bea
\lambda_{2n}(q^{\mu},...,q^{\mu}) &=&
\frac{2g^{2n}}{(4\pi)^{D/2}}\int d F_{2n-1}
\l(\frac{1}{f(x_1,x_2,...,x_{2n-1})}\r)^{n-\frac{D}{2}}
\Bigg[\frac{\Gamma\l(n-\frac{D}{2}\r)\Gamma\l(n+\frac{D}{2}\r)}
{\Gamma\l(\frac{D}{2}\r)\Gamma\l(2n\r)} \nn \\
&&+\frac{\Gamma\l(2n-\frac{D}{2}\r)}{\Gamma\l(2n\r)} +
\sum_{j=1}^{n-1}Y(D,j)\frac{\Gamma\l(2n-j-\frac{D}{2}\r)
\Gamma\l(j+\frac{D}{2}\r)}{\Gamma\l(2n-\frac{D}{2}\r)
\Gamma\l(2n\r)}\Bigg]\l(\frac{1}{q^2}\r)^{n-\frac{D}{2}}.\eea In
this expression, $\int d F_{2n-1}$ is an integration measure defined
as $$\int d F_{2n-1} = \int_0^1
\int_0^{1-x_1}...\int_0^{1-x_1-...-x_{2n-2}} dx_1
dx_2...dx_{2n-1},$$ and $$f(x_1,x_2,...,x_{2n-1}) =
\sum_{j=1}^{2n-1}j^2 x_j - \Big(\sum_{j=1}^{2n-1}x_j\Big)^2.$$
$Y(D,j)$ is a constant and determined only by the space-time
dimensionality $D$ and the value of $j$. Generically, the above
integral cannot be zero for all values of $n$. At least, it is easy
to verify that for small values of $n$, such as $n=2, 3, 4$, this
integral is nonzero. We find that the Gamma function contains no
divergence for an odd $D$, and that the measure integration is
simply equal to some constant, which leads to
$$\lambda_{2n}(q^{\mu},...,q^{\mu})\propto
g^{2n}\l(\frac{1}{q^2}\r)^{n-\frac{D}{2}}.$$

If $D$ is even, the Gamma function will be divergent for all $n \leq
D/2$. As we have explained in the calculation of
$\lambda_4(q^{\mu},q^{\mu},q^{\mu})$ in the case of $D=4$, the
existence of such divergence signals the appearance of UV divergence
in the integration over $q$. This divergence is logarithmic when $n
= D/2$, but becomes linear or even more severe if $n < D/2$. We will
also introduce a UV cutoff $\Lambda$ to eliminate such divergence.
After doing so, we obtain \bea
\lambda_{2n}(q^{\mu},...,q^{\mu})\propto \left\{
    \begin{array}{ll}
    g^{2n}\l(\frac{1}{q^2}\r)^{n-\frac{D}{2}}, & \hbox{n$\neq$D/2}
    \\
    g^{2n}\ln\l(q^2/\Lambda^2\r), & \hbox{n=D/2}
    \end{array}
    \right.
\eea In the analysis of scaling behavior of the self-coupling terms,
the function $\ln\l(q^2/\Lambda^2\r)$ makes no contribution and can
be simply replaced by a dimensionless constant, which means that
\begin{eqnarray}
\lambda_{2n}(q^{\mu},...,q^{\mu})\propto g^{2n}
\l(\frac{1}{q^2}\r)^{n-\frac{D}{2}} \quad \mathrm{as} \quad
n\rightarrow D/2.
\end{eqnarray}

Summarizing the above analysis, we finally find that
\begin{eqnarray}
\lambda_{2n}(q)\propto g^{2n}\l(\frac{1}{q^2}\r)^{n-\frac{D}{2}}
\end{eqnarray}
for all values of $n$ and $D$.


\section*{REFERENCES}
\begin{thebibliography}{1}

\bibitem{wessbook}
Wess, J. $\&$ Bagger, J. {\it Supersymmetry and Supergravity}
(Princeton Univ. Press, Princeton, 1992).

\bibitem{Weinbergbook}
Weinberg, S. {\it The Quantum Theory of Fields}, Vol. 3: {\it
Supersymmetry} (Cambridge Univ. Press, Cambridge, 2000).

\bibitem{gervais71}
Gervais, J. L. $\&$ Sakita, B. Field theory interpretation of
supergauges in dual models. {\it Nucl. Phys. B} {\bf 34}, 632--639
(1971).

\bibitem{wess74}
Wess, J. $\&$ Zumino, B. Supergauge transformations in four
dimensions. {\it Nucl. Phys. B} {\bf 70}, 39--50 (1974).

\bibitem{dimopoulos81}
Dimopoulos, S. $\&$ Georgi, H. Softly broken supersymmetry and
SU(5). {\it Nucl. Phys. B} {\bf 193}, 150--162 (1981).

\bibitem{Friedan84}
Friedan, D., Qiu, Z. $\&$ Shenkar, S. Conformal invariance,
unitarity, and critical exponents in two dimensions. {\it Phys. Rev.
Lett.} {\bf 52}, 1575--1578 (1984).

\bibitem{Nayak98}
Balents, L., Fisher, M. P. A. $\&$ Nayak, C. Nodal liquid theory of
the pseudo-gap phase of high-$T_c$ superconductors. {\it Int. J.
Mod. Phys. B} {\bf 12}, 1033--1068 (1998).

\bibitem{fendley03}
Fendley, P., Schoutens, K. $\&$ de Boer, J. Lattice models with
$\textmd{N}=2$ supersymmetry. {\it Phys. Rev. Lett.} {\bf 90},
120402 (2003).

\bibitem{Lee07}
Lee, S.-S. Emergence of supersymmetry at a critical point of a
lattice model. {\it Phys. Rev. B} {\bf 76}, 075103 (2007).

\bibitem{Huijse08}
Huijse, L., Halverson, J., Fendley, P. $\&$ Schoutens, K. Charge
frustration and quantum criticality for strongly correlated
fermions. {\it Phys. Rev. Lett.} {\bf 101}, 146406 (2008).

\bibitem{Yu10}
Yu, Y. $\&$ Yang, K. Simulating the Wess-Zumino supersymmetry model
in optical lattices. {\it Phys. Rev. Lett.} {\bf 105}, 150605
(2010).

\bibitem{Roy13}
Roy, B., Juri\v{c}i\'{c}, V. $\&$ Herbut, I. F. Quantum
superconducting criticality in graphene and topological insulators.
{\it Phys. Rev. B} {\bf 87}, 041401(R) (2013).

\bibitem{Berg13}
Bauer, B., Huijse, L., Berg, E., Troyer, M. $\&$ Schoutens, K.
Supersymmetric multicritical point in a model of lattice fermions.
{\it Phys. Rev. B} {\bf 87}, 165145 (2013).

\bibitem{Ponte14}
Ponte, P. $\&$ Lee, S.-S. Emergence of supersymmetry on the surface
of three-dimensional topological insulators. {\it New J. Phys.} {\bf
16}, 013044 (2014).

\bibitem{Grover14}
Grover, T., Sheng, D.-N. $\&$ Vishwanath, A. Emergent space-time
supersymmetry at the boundary of a topological phase. {\it Science}
{\bf 344}, 280--283 (2014).

\bibitem{Huijse14}
Huijse, L., Bauer, B. $\&$ Berg, E. Emergent supersymmetry at the
Ising-Berezinskii-Kosterlitz-Thouless multicritical point. {\it
Phys. Rev. Lett.} {\bf 114}, 090404 (2015).

\bibitem{Jian15}
Jian, S.-K., Jiang, Y.-F. $\&$ Yao, H. Emergent spacetime
supersymmetry in 3D Weyl semimetals and 2D Dirac semimetals. {\it
Phys. Rev. Lett.} {\bf 114}, 237001 (2015).

\bibitem{Rahmani15}
Rahmani, A., Zhu, X.-Y., Franz, M. $\&$ Affleck, I. Emergent
supersymmetry from strongly interacting majorana zero modes. {\it
Phys. Rev. Lett.} {\bf 115}, 166401 (2015).

\bibitem{Maciejko16}
Witczak-Krempa, W. $\&$ Maciejko, J. Optical conductivity of
topological surface states with emergent supersymmetry. {\it Phys.
Rev. Lett.} {\bf 116}, 100402 (2016).

\bibitem{Zerf16}
Zerf, N., Lin, C.-H. $\&$ Maciejko, J. Superconducting quantum
criticality of topological surface states at three loops. {\it Phys.
Rev. B} {\bf 94}, 205106 (2016).

\bibitem{Li16}
Li, Z.-X., Jiang, Y.-F. $\&$ Yao, H. Edge quantum criticality and
emergent supersymmetry in topological phases. {\it Phys. Rev. Lett.}
{\bf 119}, 107202 (2017).

\bibitem{Li18}
Li, Z.-X., Vaezi, A., Mendl, C. B. $\&$ Yao, H. Numerical
observation of emergent spacetime supersymmetry at quantum
criticality. {\it Sci. Adv.} {\bf 4}, eaau1463 (2018).

\bibitem{Jian17}
Jian, S.-K., Lin, C.-H., Maciejko, J. $\&$ Yao, H. Emergence of
supersymmetric quantum electrodynamics. {\it Phys. Rev. Lett.} {\bf
118}, 166802 (2017).

\bibitem{Chadha83}
Chadha, S. $\&$ Nielsen, H. Lorentz invariance as a low energy
phenomenon. {\it Nucl. Phys. B} {\bf 217}, 125--144 (1983).

\bibitem{Roy16}
Roy, B., Juri$\breve{c}$i$\acute{c}$, V. $\&$ Herbut, I. F. Emergent
Lorentz symmetry near fermionic quantum critical points in two and
three dimensions. {\it JHEP} {\bf 1604}, 018 (2016).

\bibitem{Chubukov03}
Abanov, A., Chubukov, A. V. $\&$ Schmalian, J. Quantum-critical
theory of the spin-fermion model and its application to cuprates:
normal state analysis. {\it Adv. Phys.} {\bf 52}, 119--218 (2003).

\bibitem{Chubukov04}
Abanov, A. $\&$ Chubukov, A. V. Anomalous scaling at the quantum
critical point in itinerant antiferromagnets. {\it Phys. Rev. Lett.}
{\bf 93}, 255702 (2004).

\bibitem{ChubukovPepinRech04}
Chubukov, A. V., P\'{e}pin, C. $\&$ Rech, J. Instability of the
quantum-critical point of itinerant ferromagnets. {\it Phys. Rev.
Lett.} {\bf 92}, 147003 (2004).

\bibitem{Metlitski10}
Metlitski, M. A. $\&$ Sachdev, S. Quantum phase transitions of
metals in two spatial dimensions. I. Ising-nematic order. {\it Phys.
Rev. B} {\bf 82}, 075127 (2010).

\bibitem{Metzner15}
Holder, T. $\&$ Metzner, W. Fermion loops and improved
power-counting in two-dimensional critical metals with singular
forward scattering. {\it Phys. Rev. B} {\bf 92}, 245128 (2015).

\bibitem{Shankar94}
Shankar, R. Renormalization-group approach to interacting fermions.
{\it Rev. Mod. Phys.} {\bf 66}, 129--192 (1994).

\bibitem{Kim08}
Kim, E.-A. \emph{et al.} Theory of the nodal nematic quantum phase
transition in superconductors. {\it Phys. Rev. B} {\bf 77}, 184514
(2008).

\bibitem{Huh08}
Huh, Y. $\&$ Sachdev, S. Renormalization group theory of nematic
ordering in d-wave superconductors. {\it Phys. Rev. B} {\bf 78},
064512 (2008).

\bibitem{Sachdev}
Sachdev, S. \textit{Quantum Phase Transitions} (Cambridge Univ.
Press, Cambridge, 2011).

\bibitem{berg2007}
Berg, E. \emph{et al.} Dynamical layer decoupling in a
stripe-ordered high-$T_c$ superconductor. {\it Phys. Rev. Lett.}
{\bf 99}, 127003 (2007).

\bibitem{berg2009}
Berg, E., Fradkin, E. $\&$ Kivelson, S. A. Charge-4e
superconductivity from pair-density-wave order in certain
high-temperature superconductors. {\it Nat. Phys.} {\bf 5},
830--833 (2009).

\bibitem{Aharony1997}
Aharony, O., Hanany, A., Intriligator, K., Seiberg, N. $\&$
Strassler, M. J. Aspects of $N=2$ supersymmetric gauge theories in
three dimensions. {\it Nucl. Phys. B} {\bf 499}, 67--99 (1997).

\bibitem{LZhao16}
Zhao, L. \emph{et al.} Emergent surface superconductivity in the
topological insulator Sb$_2$Te$_3$. {\it Nat. Commun.} {\bf 6}, 8279
(2015).

\bibitem{Belitz99}
Belitz, D., Kirkpatrick, T. R. $\&$ Vojta, T. First order
transitions and multicritical points in weak itinerant ferromagnets.
{\it Phys. Rev. Lett.} {\bf 82}, 4707--4710 (1999).

\bibitem{Metlitski15}
Metlitski, M. A., Mross, D. F., Sachdev, S. $\&$ Senthil, T. Cooper
pairing in non-Fermi liquids. {\it Phys. Rev. B} {\bf 91}, 115111
(2015).

\bibitem{Lederer15}
Lederer, S., Schattner, Y., Berg, E. $\&$ Kivelson, S. Enhancement
of superconductivity near a nematic quantum critical point. {\it
Phys. Rev. Lett.} {\bf 114}, 097001 (2015).

\end{thebibliography}
\end{document}